\documentstyle[prl,aps,epsfig]{revtex}
\tighten
\begin{document}
\title{\uppercase{Entanglement swapping with PDC sources}}
\author{M. \.Zukowski and 
D. Kaszlikowski} 
\address{Instytut Fizyki Teoretycznej i Astrofizyki,
Uniwersytet Gda\'nski, PL-80-952 Gda\'nsk, Poland
}
\date{\today}
\maketitle
\abstract{
We show that the possibility 
of distinguishing between single and 
two photon detection events is not a necessary requirement for the proof 
that recent operational realization of entanglement swapping cannot
find a local realistic description. We propose
a simple modification of the experiment, which gives a richer set of
interesting phenomena.
}

\medskip

\pacs{PACS numbers: 3.65.Bz, 42.50.Dv, 89.70.+c}

\section{Introduction}
The Bell theorem is a very important statement on the essential 
features of 
quantum mechanics. It has changed the nature of the 
earlier long disputes on
the roots of probabilistic nature of quantum predictions. The answers
to the basic questions are now expected to be found
in laboratories, rather than during seminars
 on the philosophy of science.
 The performed experimental tests give a strong
support for anyone expecting violations of Bell's inequalities also in
(future) high collection efficiency experiments.  

In 1989 it was shown \cite{GHZ89} that the
premises of the Einstein-Podolsky-Rosen argument for their claim that quantum
mechanics is an incomplete theory are inconsistent when applied to entangled
systems of three or more particles. First observations of the 
characteristic 
Greenberger-Horne-Zeilinger (GHZ) correlations
were reported by Zeilinger and his group few months ago 
\cite{INNSBRUCK}.

 The technique to obtain GHZ correlations rests upon an observation that when 
a single particle from
two independent entangled pairs is detected in a manner such that it is
impossible to determine from which pair the single came, the remaining three
particles become entangled.  This method could only be developed with clear
operational understanding of the necessary requirements to observe
multi particle interference effects for particles coming from independent
sources.

 Until recent years it was commonly believed that
particles producing EPR-Bell phenomena have to originate from a single
source, or at least have to interact with each other.  However,
under very special conditions, by a suitable monitoring procedure of the
emissions of the independent sources one can {\it pre-select} an ensemble of
pairs of particles, which either reveal EPR-Bell correlations, or are in an
entangled state.  
The first explicit proposal to use two independent sources of particles in a
Bell test was given by Yurke and Stoler \cite{YS}. 
However, they did not discuss the importance of very specific
operational requirements necessary to implement such schemes in real
experiments. Such conditions were studied in \cite{ZZHE} and 
\cite{ZZW}.

The method of entangling independently radiated photons, which share no
common past, \cite{ZZHE} is essentially a pre-selection procedure. The
selected registration acts of the idler photons define the ensemble which
contains entangled signal photons (see next sections).  Surprisingly, such a
procedure enables one to realize the Bell's idea of "event-ready" detection.
This approach for many years was thought to be completely infeasible and thus
no research was being done in that direction \cite{CH}.   This so-called 
{\it entanglement swapping} technique \cite{ZZHE}, was
also  adopted to observe
experimental quantum states teleportation \cite{BOUW}.

The first entanglement swapping experiment was 
performed in 1998 \cite{PAN}, \cite{BZ}. High visibility  (around $65\%$) two 
particle 
interference fringes were observed on a pre-selected subset of photons
that never interacted.
This is very close to the usual threshold visibility of 
two particle fringes to violate some Bell inequalities, which 
is $70.7\%$. Therefore there exists a strong temptation for
breaking this limit, and in this way showing that the two particle fringes 
due to entanglement swapping have no local and realistic model. 

However, due to the spontaneous nature of the sources involved,
the initial condition for entanglement swapping cannot
be prepared. Simply the probability that the two sources 
would produce a pair of entangled states each is of the same order as the 
probability that
one of them produces two entangled pairs.
In the latter case no entanglement swapping results. Nevertheless,
 such events can excite the trigger detectors (which in the 
case of the right initial condition select the antisymmetric 
Bell state of the two independent idlers). Therefore they are an unavoidable 
feature of the experiment, and have to be taken into account in 
any analysis of the possibility of finding a local realistic
description for the experiment.

The aim of this paper is to perform such an analysis.
We shall show that if all firings of the trigger detectors are
accepted as pre-selecting the events for a Bell-type test,
one must necessarily, at least partially, be able to 
distinguish between two and single photon events
at the detectors observing the signals to enable demonstrations
 of violations of local realism. Whereas, if one accepts additional 
selection at the trigger detectors, based on the polarisation 
of the idlers,  detectors possessing this ability are unnecessary.
We shall present our argumentation assuming that reader
knows the methods and results of \cite{ZZHE}, \cite{ZZW} and \cite{PAN}.
The analysis will be confined to the gedanken situation
of perfect detection efficiency (the results can be easily generalised
to the non-ideal case).
\begin{figure}
\centerline{\psfig{width=5.0cm,angle=-90,file=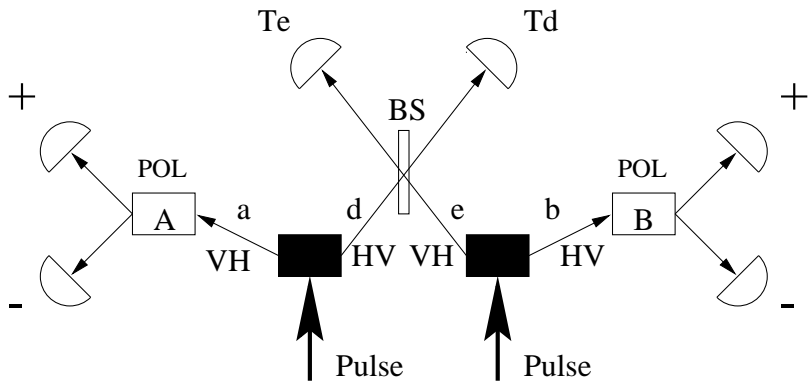}}
\vspace{2cm} 
\caption{
Entanglement swapping. Two type-II 
down conversion crystals are pumped
by a pulsed laser. The radiation from each of the crystals   
is entangled in polarisations (e.g. if one has an H polarised photon in 
mode $a$, then in mode $b$ is a V polarised photon). The idler photons
(in modes $d$
and $e$) are fed into a non-polarising beam splitter BS.
Simultaneous firing of the trigger detectors $T_e$ and $T_d$
pre-selects, in the event-ready way, a sub-ensemble of detection events
behind the polarising beam splitters $A$ and $B$.
The orientation of the 
polarising beam splitters can be set at will by the 
local observers. The output 
signal photons 
are registered by the two local detection stations, consisting of 
detectors denoted by $+$ and $-$. }
\end{figure}

\section{Entanglement swapping and local realism}
Consider the set-up of fig. 1, which is in 
principle the scheme used in the Innsbruck experiment
\cite{PAN}.
Two pulsed type-II 
down conversion sources are 
emitting their radiation into the 
spatial propagation modes $a$ and $b$ (signals), $c$ and $d$ (idlers).
Due to
the statistical properties of the PDC radiation,
the initial state that is fed to the 
interferometric set-up has the following form:
\begin{eqnarray}
&|\psi\rangle=\sum_{n=0}^{\infty}({\gamma\over \sqrt2}
(a_{H}^{\dagger}d_{V}^{\dagger}+a_{V}^{\dagger}
d_{H}^{\dagger}))^{n}&\nonumber\\
&\times\sum_{m=0}^{\infty}({\gamma\over \sqrt2}
(e_{H}^{\dagger}b_{V}^{\dagger}+e_{V}^{\dagger}
b_{H}^{\dagger}))^{m}|0\rangle,&
\label{e1}
\end{eqnarray}
where, for instance, $a_{H}^{\dagger}$ denotes the creation operator of
the photon in beam $a$ having ``horizontal" polarisation.
As for the entanglement swapping to work one cannot have too excessive 
pump powers \cite{ZZHW},
the $\gamma$ coefficient can be assumed small.
Therefore we select only those  terms that are proportional 
to $\gamma^{2}$, as these are the lowest order terms
terms that can induce simultaneous firing of both trigger detectors.
They read
\begin{eqnarray}
&|\psi'\rangle={1\over 2}{\gamma}^{2}((a_{H}^{\dagger}d_{V}^{\dagger}+
d_{V}^{\dagger}d_{H}^{\dagger})(e_{H}^{\dagger}b_{V}^{\dagger}+
e_{V}^{\dagger}b_{H}^{\dagger})&\nonumber\\
&+(a_{H}^{\dagger}d_{V}^{\dagger}+a_{V}^{\dagger}d_{H}^{\dagger})^{2}+
(e_{H}^{\dagger}b_{V}^{\dagger}+
e_{V}^{\dagger}b_{H}^{\dagger})^{2})|0\rangle.&
\label{e2}
\end{eqnarray}
The factor 
$\frac{1}{2}\gamma^2$ simply gives the order of magnitude of 
the probability of the two trigger detectors to fire, and therefore we
drop it from further considerations. The action of the 
non-polarising beam splitter (BS) is described by
$d_{x}^{\dagger}={1\over \sqrt2}({\tilde d}_{x}^{\dagger}+
i{\tilde e}_{x}^{\dagger})$ and
$e_{x}^{\dagger}={1\over\sqrt2}({\tilde e}_{x}^{\dagger}+
i{\tilde d}_{x}^{\dagger})$
where $x=H$ or $x=V$, and $\tilde{e}$ and $\tilde{d}$
represent the modes monitored by the trigger detectors behind
the beam splitter.  
Taking into account only the terms in (\ref{e2}) that lead to clicks
at two trigger detectors we arrive at
\begin{eqnarray}
&|\psi''\rangle=(i(a_{H}^{\dagger 2}+b_{H}^{\dagger 2})
{\tilde e_{V}^{\dagger}}{\tilde d_{V}^{\dagger}}+
i(a_{V}^{\dagger 2}+b_{V}^{\dagger 2}){\tilde e_{H}^{\dagger}}
{\tilde d_{H}^{\dagger}}&\nonumber\\
&+i(a_{H}^{\dagger}a_{V}^{\dagger}+b_{H}^{\dagger}b_{V}^{\dagger})
({\tilde e_{V}}^{\dagger}{\tilde d_{H}}^{\dagger}+
{\tilde e_{H}}^{\dagger}{\tilde d_{V}}^{\dagger})&\nonumber\\
&{1\over2}(a_{H}^{\dagger}b_{V}^{\dagger}-a_{V}^{\dagger}
b_{H}^{\dagger})({\tilde e_{V}}^{\dagger}{\tilde d_{H}}^{\dagger}-
{\tilde e_{H}}^{\dagger}{\tilde d_{V}}^{\dagger}))|0\rangle.&
\label{e4}
\end{eqnarray}
It is convenient to normalise and rewrite the above state into the form:
\begin{eqnarray}
&|\psi_{N}\rangle={1\over\sqrt13}
\left[i\sqrt2({1\over\sqrt2}a_{H}^{\dagger 2}+
{1\over\sqrt2}b_{H}^{\dagger 2})
|VV\rangle\right.&\nonumber\\
&\left.+i\sqrt2({1\over\sqrt2}a_{V}^{\dagger 2}+
{1\over\sqrt2}b_{V}^{\dagger 2})|HH\rangle\right.&\nonumber\\
&\left.+(i(a_{H}^{\dagger}a_{V}^{\dagger}+
b_{H}^{\dagger}b_{V}^{\dagger})+
{1\over2}(a_{H}^{\dagger}b_{V}^{\dagger}-a_{V}^{\dagger}
b_{H}^{\dagger}))|VH\rangle\right.&\nonumber\\
&\left.+(i(a_{H}^{\dagger}a_{V}^{\dagger}+
b_{H}^{\dagger}b_{V}^{\dagger})-
{1\over2}(a_{H}^{\dagger}b_{V}^{\dagger}-a_{V}^{\dagger}
b_{H}^{\dagger})\right]|HV\rangle,&
\label{e4a}
\end{eqnarray}
where $|VV\rangle={\tilde 
e_{V}^{\dagger}}{\tilde d_{V}^{\dagger}}|0\rangle$, 
$|HH\rangle=
{\tilde e_{H}^{\dagger}}
{\tilde d_{H}^{\dagger}}|0\rangle$, 
$|VH\rangle=
{\tilde e_{V}}^{\dagger}{\tilde d_{H}}^{\dagger}|0\rangle$ and 
$|HV\rangle=
{\tilde e_{H}}^{\dagger}{\tilde d_{V}}^{\dagger}|0\rangle$.
We see  clearly that several processes may lead to 
the simultaneous firing of the trigger detectors (which observe 
the spatial modes) 
$\tilde{e}$ and $\tilde{d}$. 
The signal photons enter the polarising 
beam splitters.  Their action can be
described by the following relations
\begin{eqnarray}
&x_{V}^{\dagger}=\cos(\theta_{i})x_{+}^{\dagger}+
\sin(\theta_{i})x_{-}^{\dagger}&\nonumber\\
&x_{H}^{\dagger}=-\sin(\theta_{i})x_{+}^{\dagger}+
\cos(\theta_{i})x_{-}^{\dagger},&
\end{eqnarray}
with $x=a,b; i=1,2$ respectively and $+$, $-$ 
denoting the output spatial modes.

The probabilities of various
two-particle processes
that may at the spatially separated observation stations, under the
condition of both trigger detectors firing simultaneously, are given by:
\begin{eqnarray}
&P(1a_{+},1a_{-};0b_{+},0b_{-})=P(2a_{+},0a_{-};0b_{+},0b_{-})&\nonumber\\
&=P(0a_{+},2a_{-};0b_{+},0b_{-})=P(0a_{+},0a_{-};1b_{+},1b_{-})&\nonumber\\
&=P(0a_{+},0a_{-};2b_{+},0b_{-})=P(0a_{+},0a_{-};0b_{+},2b_{-})=
{2\over13},&\nonumber\\
&P(1a_{+},0a_{-};1b_{+},0b_{-})=P(0a_{+},1a_{-};0b_{+},1b_{-})
={1\over26}\sin(\theta_{1}-\theta_{2})^2,&\nonumber\\
&P(1a_{+},0a_{-};0b_{+},1b_{-})=P(0a_{+},1a_{-};1b_{+},0b_{-})
={1\over26}\cos(\theta_{1}-\theta_{2})^2,&
\label{e5}
\end{eqnarray}
where, for example, $P(0a_{+},0a_{-};2b_{+},0b_{-})$  denotes
the probability of observing two photons at the output $b_{+}$, and no 
photons in the other outputs.

The Bell correlation function 
for the product of the measurement results
on the signals at the two sides of the experiment
can be redefined in the way proposed in \cite{POP}.
All standard Bell-type events are assigned their usual values. 
For example, if there is a single photon 
click at the detector $\pm$ at the station 
A, the assigned value is $\pm1$.
However, all non-standard events are assigned the value of one.
I.e., if no photons are registered at one side, the local value of the 
measurement is one, if two photons are registered at one side again
the local measurement value
 is one. The latter case includes both the event in which the two photons 
end-up at a single detector, as well as those when two detectors 
at the local station fire. Please note, that the experiment considered 
is a realization of Bell's idea of 
``event ready detectors'' (see e.g. \cite{CH}). Therefore,
non-detection events are operationally well defined (as the simultaneous 
firing of the trigger detectors pre-selects the sub-ensemble
of time intervals in which one can expect the signal detectors to fire).

The above value assignment method works perfectly if one 
assumes that it is possible to
distinguish between single and double photon detection at a single 
detector. However, this is 
usually not
the case. Thus it is convenient to have a parameter $\alpha$ that
measures the distinguishability of the double and single detection at
one detector ($0\leq\alpha\leq1$, and gives
the probability of distinguishing, by the employed devices,
 of the double counts). 

The partial distinguishability blurs the 
distinction
between events (at one side) in which 
there was a one photon detected at say the output $\pm$, and 
events in which two photons entered a the detector observing output
$\pm$, but the detector failed to distinguish this event
from a single photon count.
In such a case the local event is sometimes ascribed by the local
observer a wrong value
namely $-1$ instead of $1$ (if both photons
go to the $`` - "$ exit of the polariser
and the devices fail to inform the experimenter 
that it is a two photon event, this is interpreted as a firing due 
to a single photon and is ascribed a $-1$ value). 
Please note, that if one includes less than perfect detection efficiency
of the detectors this problem is more frequent and more involved
(we shall not study this aspect here).

Under such a value assignment the correlation function reads:
\begin{eqnarray}
&E_{\alpha}(\theta_{1},\theta_{2})=-{1\over 13}\cos(2\theta_{1}-2\theta_{2})+
{4\over 13}(1+2\alpha),&
\end{eqnarray}
where $\alpha$ is the numerical value of the distinguishability.
When we put into the standard CHSH inequality this correlation
function it violates the standard bound of 2, only if the distinguishability
satisfies $\alpha\geq{9-\sqrt2\over8}\approx0.948$. 
Such values are definitely
beyond the current technological limits. As the efficiency
of real detectors makes this problem even more acute, one has to
propose a modification of the experiment that gets rid of this problem.

Therefore,
in front of the idler detector $Te$ we propose to put 
polarising beam splitter that transmit only vertical polarisation
whereas in front of the idler detector $Td$ one that transmits only
horizontal polarisation. This further reduces the 
relevant terms in our state, i.e. those that can
induce firing of the trigger detectors, to the following 
ones:

\begin{eqnarray}
&|\psi\rangle=\sqrt{2\over5}\left(i(a_{H}^{\dagger}a_{V}^{\dagger}+
b_{V}^{\dagger}b_{H}^{\dagger})+
{1\over2}(a_{H}^{\dagger}b_{V}^{\dagger}-
a_{V}^{\dagger}b_{H}^{\dagger})\right){\tilde e}_{V}^{\dagger}
{\tilde d}_{H}^{\dagger}|0\rangle.&
\label{e5a}
\end{eqnarray}
Again we have normalised the above state.

Using the above formula we can calculate the probabilities of all possible
processes in this interferometric set-up, conditional on
 firings of the two trigger detectors:

\begin{eqnarray}
&P(1a_{+},1a_{-};0b_{+},0b_{-})={2\over5}\cos(2\theta_{1})^2,&\nonumber\\
&P(2a_{+},0a_{-};0b_{+},0b_{-})=P(0a_{+},2a_{-};0b_{+},0b_{-})
={1\over5}\sin(2\theta_{1})^2,&\nonumber\\
&P(0a_{+},0a_{-};1b_{+},1b_{-})={2\over5}\cos(2\theta_{2})^2,&\nonumber\\
&P(0a_{+},0a_{-};2b_{+},0b_{-})=P(0a_{+},0a_{-};0b_{+},2b_{-})
={1\over5}\sin(2\theta_{2})^2,&\nonumber\\
&P(1a_{+},0a_{-};1b_{+},0b_{-})
=P(0a_{+},1a_{-};0b_{+},1b_{-})={1\over10}\sin(\theta_{1}-\theta_{2})^2,&
\nonumber\\
&P(1a_{+},0a_{-};0b_{+},1b_{-})=P(0a_{+},1a_{-};1b_{+},0b_{-})
={1\over10}\cos(\theta_{1}-\theta_{2})^2.&
\label{e7}
\end{eqnarray}

Under the earlier defined value assignment the correlation function
for the current version of the experiment reads:
\begin{eqnarray}
&E_{\alpha}(\theta_{1},\theta_{2})=-{1\over5}\cos(2\theta_{1}-2\theta_{2})&
\nonumber\\
&+{2\over5}(1-\alpha)((\cos2\theta_{1})^{2}+(\cos2\theta_{2})^2)
+{4\over5}\alpha.&
\end{eqnarray}
When such a correlation functions are inserted into the CHSH inequality
one has:
\begin{eqnarray}
&-2\leq-{1\over5}[\cos2(\theta_{1}-\theta_{2})+\cos2(\theta_{1}-\theta_{2}')&
\nonumber\\
&+\cos2(\theta_{1}'-\theta_{2})-\cos2(\theta_{1}'-\theta_{2}')]&\nonumber\\
&+{4\over5}(1-\alpha)((\cos2\theta_{1})^2+(\cos2\theta_{2})^2)+
{8\over5}\alpha\leq 2.&
\label{CHSH}
\end{eqnarray}
Please note that some of the terms of the correlation function 
which depend only on one local angle cancel upon insertion into CHSH
inequality.

For $\alpha=1$ (perfect distinguishability) the 
middle expression in (\ref{CHSH}) reaches
2.16569, i.e. we have a clear violation 
of the local realistic bound.
This maximal violation occurs at angles (in radians)
 $2\theta_{1}=-1.30278,$ $2\theta_{1}'=-2.87435,$ $2\theta_{2}=1.05326,$
$2\theta_{2}'=2.62386$.
What is more interesting, for $\alpha=0$, i.e. for a complete lack of
distinguishability
between two and single photon events at one detector, the
expression in (\ref{CHSH})
reaches a value which is not much lower, namely 
2.11453. This can be reached for the orientation angles 
$2\theta_{1}=0.0837317,$ $2\theta_{1}'=-1.0749,$ 
$2\theta_{2}=3.05769,$ $2\theta_{2}'=4.21568$). 

Therefore we conclude
that the proposed modification of the entanglement swapping experiment 
makes possible, despite the unwanted additional events
due to the impossibility of controlling the spontaneous emissions
 at the two separate sources, makes it possible to consider it
as test of Bell inequalities. The standard configuration can serve
as a test of local realism only under the condition of
extremely high distinguishability between two and single photon counts.

Finally let us mention that the proposed
modification in the entanglement swapping configuration
enables one to observe,  in the event ready mode,
a bosonic interference effect similar to the so-called 
Hong-Ou-Mandel dip \cite{HOM}. It is described by the first four formulas of
(\ref{e7}). E.g. if $\theta_1=\pi/4$, 
no coincidences between firings of the two detectors
of the station $a$
are allowed. All two photon events at this station
are, under this setting, double counts at a single detector.
Thus, we have two very interesting non-classical 
phenomena in one experiment.

\medskip
\noindent {\bf Acknowledgments}
MZ was supported by the University of Gdansk Grant No
BW/5400-5-0264-9. DK was supported by the KBN Grant 2 P03B 096 15.
MZ thanks Anton Zeilinger and Harald Weinfurter for years of 
discussions on the subject, and for the organisers of the Workshop
for hospitality.

\end{document}